\documentclass{PoS}
\usepackage{amssymb,amsmath}

\def\eps{\varepsilon}
\def\epe{\varepsilon'/\varepsilon}

\title{Quo vadis flavour physics?\\
{\large FPCP2017 theory summary and outlook}}

\ShortTitle{Quo vadis flavour physics?}

\author{\speaker{Monika Blanke}\\
        {Institut f\"ur Kernphysik, Karlsruhe Institute of Technology,
  Hermann-von-Helmholtz-Platz 1,
  D-76344 Eggenstein-Leopoldshafen, Germany}\\
 {Institut f\"ur Theoretische Teilchenphysik,
  Karlsruhe Institute of Technology, Engesserstra\ss e 7,
  D-76128 Karlsruhe, Germany} \\
        E-mail: \email{monika.blanke@kit.edu}}

\abstract{We review the recent highlights of theoretical flavour physics, based on the theory summary talk given at FPCP2017. Over the past years, a number of intriguing anomalies have emerged in flavour violating $K$ and $B$ meson decays, constituting some of the most promising hints for the presence of physics beyond the Standard Model. We discuss the theory status of these anomalies and outline possible future directions to test the underlying New Physics.}

\FullConference{The 15th International Conference on Flavor Physics \& CP Violation\\
		5-9 June 2017\\
		Prague, Czech Republic}

\begin{document}

\section{New Physics sensitivity of flavour physics}\label{sec:intro}

Despite the impressive amount of data collected and analysed during run 1 and early run 2 of the LHC, until now no clear signal of new particles has been found in the direct searches for New Physics (NP). While we might have simply been unlucky in the choice of NP discovery modes, it is also conceivable that the NP scale is larger than the energies directly accessible at the LHC.  Together with the continuing efforts at the high-$p_T$ frontier, it is therefore of utmost importance to explore alternative routes to access the nature of NP. 

Precision tests of Standard Model (SM) observables provide complementary access to NP contributions. As in such low energy processes new particles contribute via quantum effects, they are not plagued by the same strict energy cut-off as the LHC. The sensitivity in this case is limited by the cleanliness of the theoretical predictions in the SM, in addition to the obtained experimental precision. Particularly useful in this context are observables whose SM contribution is suppressed, increasing their potential NP sensitivity. 

Flavour and CP-violating observables hence play a key role in the hunt for NP. In the SM, flavour changing neutral current processes are strongly suppressed not only by a loop factor, but also by the smallness of the off-diagonal CKM elements and the GIM mechanism. NP contributions, even if generated well beyond the TeV scale, can therefore be significant, provided the theoretical and experiment precision is sufficient. The highest NP sensitivity is obtained in rare kaon decays due to their strong CKM suppression by $V_{ts}^*V_{td}$ and their theoretical cleanliness, in particular in the $K\to\pi\nu\bar\nu$ decays. Indeed the latter have been shown to have the potential to probe NP scales beyond $1000\,\text{TeV}$ \cite{Buras:2014zga}. $B$ decays on the other hand, while being less sensitive to extremely high scales, have the advantage to offer a multitude of observables suitable to disentangle the underlying NP structure at work.

In deciphering the NP flavour structure, the study of correlations between flavour and CP-violating observables will be a crucial task. In this context, we can distinguish between two different types of correlations: 
\begin{itemize}
\item Correlations between observables within a given meson system give information on the underlying NP
operator structure. In this way, contributions from vector or scalar mediators can be distinguished, or the chirality of the NP coupling to the SM quarks and leptons can be determined.
\item
Correlations between related observables in different meson systems, on the other hand, allow to draw
conclusions on the underlying flavour symmetry. For example, specific patterns of effects are predicted in models with Minimal Flavour Violation \cite{Buras:2000dm,DAmbrosio:2002vsn,Buras:2003jf} or with a minimally broken $U(2)^3$ flavour symmetry \cite{Barbieri:2011ci,Barbieri:2012uh,Buras:2012sd}.
\end{itemize}

The first step towards the identification of NP in the flavour sector is however to establish deviations from the SM predictions in flavour and CP-violating decays. While at the moment such an unambiguous NP observation is still outstanding, a number of intriguing anomalies have emerged in the field of flavour physics. In the remainder of this contribution, we focus on the ones that recently received the most attention both in and beyond the flavour physics community, dedicating a sectiion to each of them. These are:
\begin{enumerate}
\item the tension between the measured amount of direct CP violation in $K\to\pi\pi$ decays and its SM prediction, as recently obtained from lattice QCD determinations \cite{Bai:2015nea} and confirmed by dual QCD calculations \cite{Buras:2015xba,Buras:2016fys},
\item the persisting signs of lepton flavour universality violation seen by several experimental collaborations in the semileptonic $B\to D^{(*)}\ell\nu$ decays \cite{Amhis:2016xyh},
\item and last but not least the anomalies in $b\to s\mu^+\mu^-$ transitions like $B\to K^{(*)}\mu^+\mu^-$ and related lepton flavour universality ratios pointed out by the LHCb collaboration \cite{Aaij:2015oid,Aaij:2017vbb}.
\end{enumerate}

In passing we note that there are other unresolved puzzles related to the flavour sector, like the anomalous magnetic moment of the muon \cite{Hagiwara:2017zod} or the $B_{s,d}$ meson oscillation frequency \cite{Bazavov:2016nty,Blanke:2016bhf}.

\section{Direct CP violation in kaon decays}

The theoretical description of direct CP violation in $K\to\pi\pi$ decays has been a long-standing problem. A precise experimental value \cite{Batley:2002gn,AlaviHarati:2002ye,Abouzaid:2010ny},
\begin{equation}
\text{Re}(\epe)_\text{exp} = (16.6\pm 2.3)\cdot 10^{-4}\,,
\end{equation}
existed since the early 2000s. However it took until 2015 until a first SM calculation became available. A big step forward has been made by the RBC-UKQCD collaboration \cite{Bai:2015nea} who presented the first lattice calculation of the hadronic matrix elements
\begin{equation}\label{eq:B6B8}
B_6^{(1/2)} = 0.57\pm 0.19\,, \qquad B_8^{(3/2)} = 0.76\pm 0.05\,,
\end{equation}
resulting in a SM value for $\epe$ that is significantly lower than the measured value. 

The strict large $N_c$ limit predicts $B_6^{(1/2)}$ and $B_8^{(3/2)}$ to be equal to unity. Yet a recent analysis within the dual QCD approach shows a suppression of both matrix elements below this naive value, resulting in the bound \cite{Buras:2015xba,Buras:2016fys}
\begin{equation}\label{eq:dual}
B_6^{(1/2)} < B_8^{(3/2)} < 1\,.
\end{equation}
Interestingly \eqref{eq:dual} not only supports the RBC-UKQCD result \eqref{eq:B6B8}, but also unambiguously predicts the presence of NP in $\epe$. These results stimulated a number of phenomenological studies, both in and beyond the SM.

In the SM, a simple phenomenological expression for $\epe$ has been derived \cite{Buras:2003zz,Buras:2015yba}:
\begin{equation}
\text{Re}(\epe)=\frac{\text{Im}(V_{ts}^* V^{}_{td})}{1.4\cdot 10^{-4}}\cdot 10^{-4} \cdot\left[\left(
-3.6+21.4 \,B_6^{(1/2)} \right) + \left( 1.2-10.4 \, B_8^{(3/2)} \right) \right] \,.
\end{equation}
Here, the first term in the brackets represents the $\Delta I =1/2$ amplitude that is mainly generated by QCD penguin contributions. The second term corresponds to the $\Delta I =3/2$ contribution, mainly due to electroweak penguins. The two contributions largely cancel each other, resulting in a very small SM prediction. 

Two independent phenomenological analyses have determined the SM prediction at next-to-leading order, taking into account the RBC-UKQCD result \cite{Bai:2015nea}, with consistent results:
\begin{equation}
\text{Re}(\epe)_\text{SM} = (1.9\pm4.5)\cdot 10^{-4} \quad\text{\cite{Buras:2015yba}}\,,\qquad
\text{Re}(\epe)_\text{SM} = (1.06\pm5.07)\cdot 10^{-4} \quad\text{\cite{Kitahara:2016nld}}\,.
\end{equation}
These numbers are almost $3\sigma$ below the data. The theoretically challenging next-to-next-to-leading order calculation is currently in progress \cite{Cerda-Sevilla:2016yzo}.

\begin{figure}
\centering{\includegraphics[width=.65\textwidth]{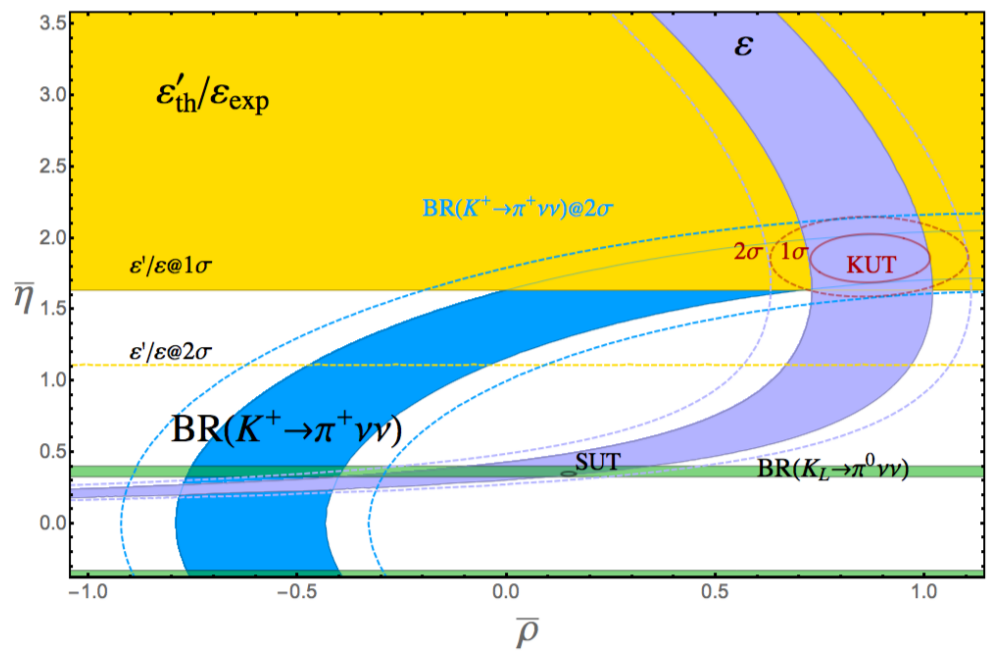}}
\caption{$K$-unitarity triangle, assuming today's central values (SM prediction for $K_L\to\pi^0\nu\bar\nu$) and uncertainties expected to be reached by 2025. Figure taken from \cite{Lehner:2015jga}.\label{fig:KUT}} 
\end{figure}

With the measurement of a number of theoretically well-understood kaon decay observables, one can construct the unitarity triangle from $K$ physics only ($K$-unitarity triangle) \cite{Buras:1994ec,Buchalla:1994tr,Lehner:2015jga}. Besides $\eps_K$, which is already known to be an important player in global unitarity triangle fits \cite{Charles:2015gya,Bevan:2014cya}, crucial imput will be provided by the branching ratios of the extremely clean decays $K^+\to\pi^+\nu\bar\nu$ and $K_L\to\pi^0\nu\bar\nu$. Finally, with improved lattice determinations of the hadronic matrix elements $B_6^{(1/2)}$ and $B_8^{(3/2)}$, $\epe$ will overconstrain the $K$-unitarity triangle. A future mismatch in the $K$-unitarity triangle constraints, as indicated in figure \ref{fig:KUT}, would be a definite signal of NP contributions to the kaon sector.

As discussed in section \ref{sec:intro}, the strong suppression of $\epe$ in the SM paves the way to possible large NP contributions. It is therefore not surprising that the recent SM calculations indicate a tension with the experimental value. Indeed, theoretical studies have been performed in a number of NP models,  with the outcome that in many NP scenarios it is possible to enhance $\epe$ by an order to magnitude and resolve the tension. We refer the reader to the original publications for further details on Little Higgs models \cite{Blanke:2015wba},  supersymmetric models \cite{Tanimoto:2016yfy,Endo:2016aws,Kitahara:2016otd,Crivellin:2017gks}, simplified models with flavour changing $Z$ or $Z'$ couplings \cite{Buras:2015yca,Endo:2016tnu},  331 models \cite{Buras:2016dxz}, vector-like quark models \cite{Bobeth:2016llm}, and a model independent analysis \cite{Buras:2015jaq}. A review can be found e.\,g.\ in \cite{Blanke:2016uwb}.

\begin{figure}
\centering{\includegraphics[width=.95\textwidth]{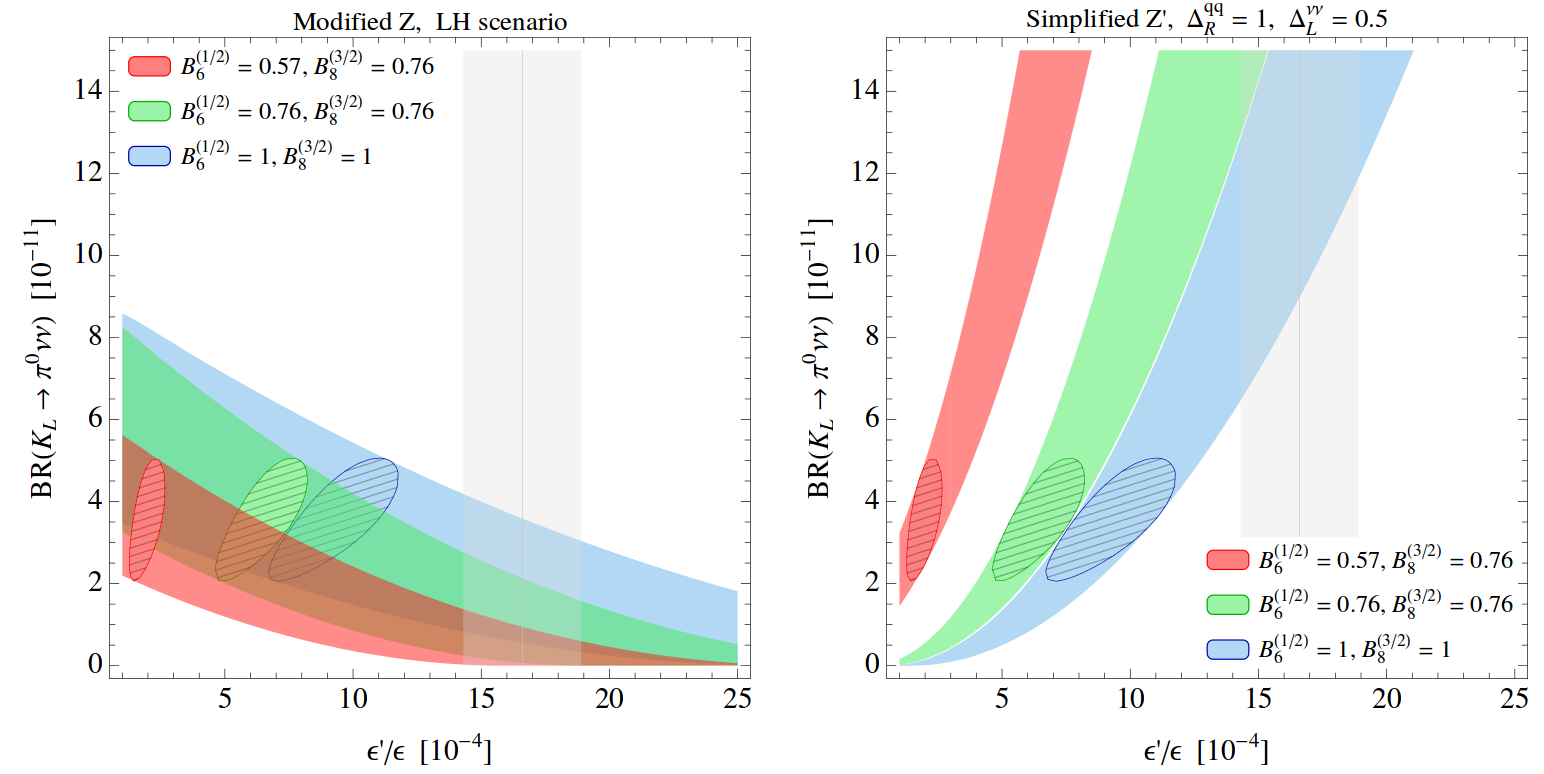}}
\caption{Correlation between $\varepsilon'/\varepsilon$ and $\mathcal{B}(K_L\to\pi^0\nu\bar\nu)$ in simplified models with tree level flavour changing $Z$ and $Z'$ couplings.
Figure taken from \cite{Buras:2015yca}.\label{fig:epeKL}} 
\end{figure}

Once the presence of NP in $\epe$ has been established, the task will be to disentangle its structure. To this end it will be necessary to precisely  determine the NP contributions to other rare kaon decay observables, like the branching ratios of $K_L\to\pi^0\nu\bar\nu$ and $K^+\to\pi^+\nu\bar\nu$. The latter play a unique role in flavour physics, due to their outstanding theoretical cleanliness both in and beyond the SM. As both $\epe$ and $\mathcal{B}(K_L\to\pi^0\nu\bar\nu)$ measure direct CP violation in kaon decays, the deviations from the SM in both observables are correlated. Interestingly, the correlation can either be direct, meaning that both observables are simultaneously enhanced, or reciprocal, meaning that the enhancement of one observable goes along with a suppression of the other. The nature of the correlation depends on the NP coupling structure at work, as has been shown in \cite{Buras:2015yca} for different simplified models with tree level flavour changing $Z$ and $Z'$ couplings. The result is displayed in figure \ref{fig:epeKL}

To further understand the NP structure at work, it is then useful to consider also the correlation between  $K_L\to\pi^0\nu\bar\nu$ and $K^+\to\pi^+\nu\bar\nu$. While the latter mode is CP-conserving, its correlation with the CP-violating $K_L$ decay provides insight on the operator structure at work in neutral kaon mixing. In models with only left-handed flavour changing transitions, the strong constraint from the CP-violating parameter $\eps_K$ implies a NP phase in the $s\to d\nu\bar\nu$ that is close to a multiple of $\pi/2$, in turn implying a striking correlation in the $K\to\pi\nu\bar\nu$ plane \cite{Blanke:2009pq}.

In order to fully exploit the NP discovery potential of kaon physics, it is of utmost important to improve the SM predictions as much as possible. More precise lattice determinations of the $K\to\pi\pi$ matrix elements as well as an independent confirmation of the obtained results are required in order to establish the presence of NP in $\epe$. Lattice QCD however also plays an important role in other kaon decay observables, and a calculation of the long-distance contributions to the mass difference $\Delta M_K$ \cite{Bai:2014cva} and the decay $K^+\to\pi^+\nu\bar\nu$ \cite{Bai:2017fkh} will be extremely valuable. But equally important is a precise knowledge of the parameters of the CKM matrix, mainly $|V_{cb}|$ and the angle $\gamma$, to which kaon observables are highly sensitive. Further progress, both on the theory and on the experimental side, is therefore eagerly awaited.

\section{\boldmath Lepton flavour universality and semitauonic $B$ decays}

Meson decays mediated by charged current interactions are generally expected to be less sensitive to NP contributions, as they already arise at the tree level in the SM. An exception to this rule of thumb is however provided by observables that are theoretically very clean, so that even relatively small NP contributions become apparent. This is the case for the ratios
\begin{equation}\label{eq:RD}
R(D^{(*)}) = \frac{\mathcal{B}(B\to D^{(*)}\tau\nu)}{\mathcal{B}(B\to D^{(*)}\ell\nu)}\,,
\end{equation}
that constite a test of lepton flavour universality (LFU). 

In the SM, LFU is violated by the difference in the charged lepton masses: $m_\tau \gg m_e,m_\mu$.
Consequently, the SM predictions \cite{Bernlochner:2017jka,Bigi:2017jbd} for $R(D)$ and $R(D^*)$ differ from unity:
\begin{equation}\label{eq:RDSM}
R(D)_\text{SM} = 0.299 \pm 0.003\,,\qquad
R(D^*)_\text{SM} = 0.257 \pm 0.003\,.
\end{equation}
As hadronic uncertainties largely cancel in the ratio, a high precision in the SM prediction can be reached.

Over the past years, BaBar \cite{Lees:2013uzd}, Belle \cite{Huschle:2015rga,Sato:2016svk,Hirose:2016wfn} and LHCb \cite{Aaij:2015yra,Wormser} presented a number of measurements of these ratios, yielding a consistent HFLAV fit \cite{Amhis:2016xyh}
\begin{equation}\label{eq:RDexp}
R(D)_\text{exp} = 0.407 \pm 0.039 \pm 0.024\,, \qquad
R(D^*)_\text{exp} = 0.304 \pm 0.013 \pm 0.007\,.
\end{equation} 
The data thus exhibit a $4.1\sigma$ deviation from the SM, thereby hinting for new sources of LFU violation.

Model-independently, the relevant $b\to c \tau\nu$ transition can be described by the effective Hamiltonian
\begin{equation}\label{eq:bctaunu}
\mathcal{H}_\text{eff}^{b\to c\tau\nu} = \frac{4 G_F}{\sqrt{2}} V_{cb} \mathcal{O}_{V_L}+\frac{1}{\Lambda^2} \sum_j C_j \mathcal{O}_j + h.c.\,,
\end{equation}
with $\Lambda$ denoting the NP scale, and the four-fermion operators
\begin{eqnarray}
\mathcal{O}_{V_{L,R}} &=& (\bar c\gamma^\mu P_{L,R} b)(\bar\tau\gamma_\mu P_L \nu)\,,\\
\mathcal{O}_{S_{L,R}} &=& (\bar c P_{L,R} b)(\bar\tau P_L \nu)\,,\\
\mathcal{O}_T &=& (\bar c\sigma^{\mu\nu} P_L b)(\bar\tau\sigma_{\mu\nu}P_L \nu)\,.
\end{eqnarray}
In the SM, the  transition is mediated by a tree-level $W^\pm$ exchange, hence only the operator $\mathcal{O}_{V_L}$ is present. 

\begin{figure}
\centering{\includegraphics[width=.85\textwidth]{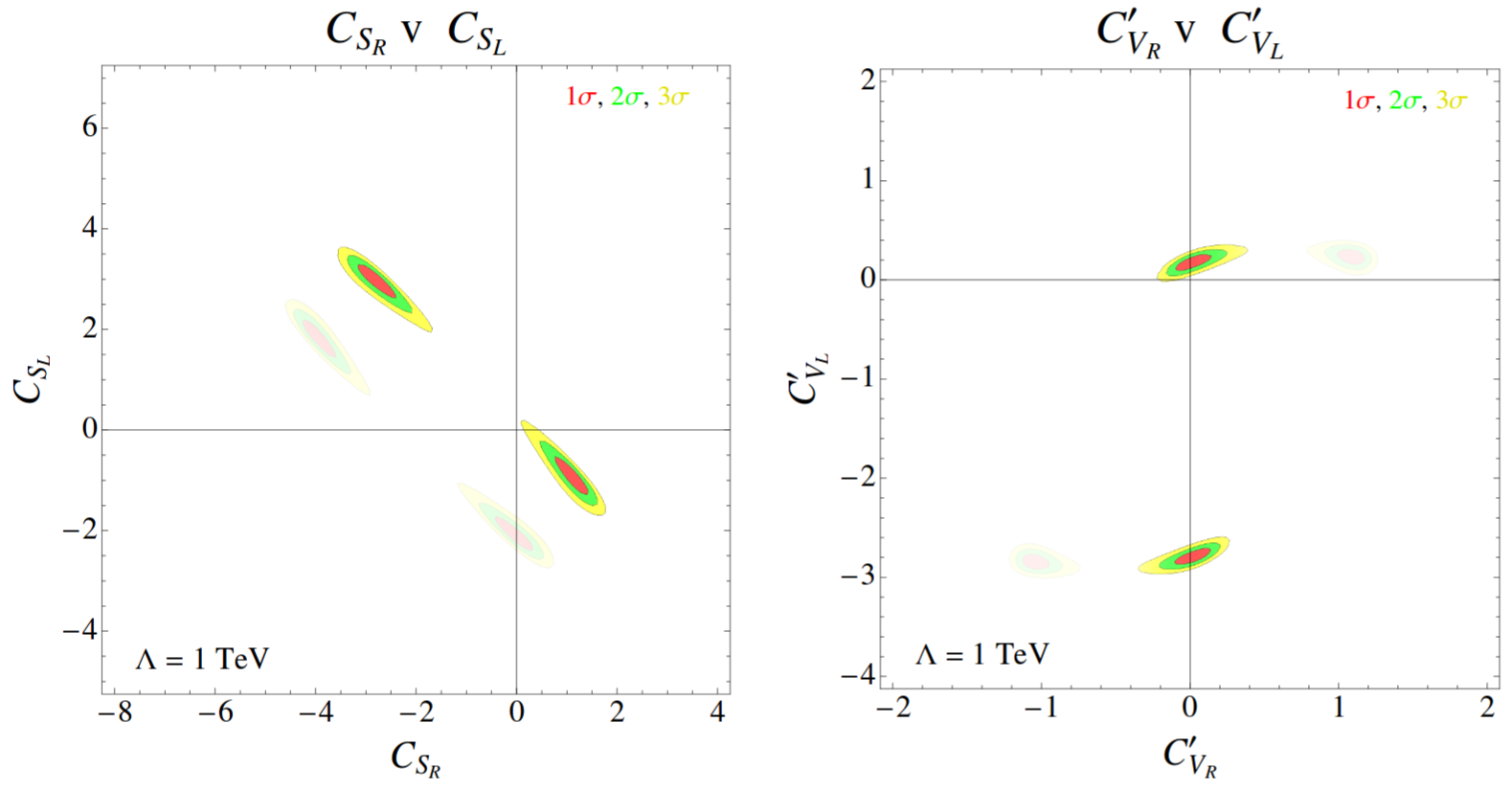}}
\caption{Model-independent fit of the effective Hamiltonian \eqref{eq:bctaunu} to the $R(D)$ and $R(D^*)$ data. 
Figure taken from \cite{Freytsis:2015qca}.\label{fig:RDfit}} 
\end{figure}

Comparing \eqref{eq:RDSM} with \eqref{eq:RDexp}, it is easy to see that the required NP contribution is quite large, having to compete with a tree-level process in the SM. Indeed, fits \cite{Fajfer:2012jt,Freytsis:2015qca,Bardhan:2016uhr} of the effective Hamiltonian \eqref{eq:bctaunu} to the data indicate that the $R(D^{(*)})$ anomaly can be resolved by the presence of NP in either the scalar or the vector Wilson coefficients, as shown in figure \ref{fig:RDfit}. In the first case, $C_{S_L}\simeq -C_{S_R}$ is necessary in order to generate NP effects of similar size in $R(D)$ and $R(D^*)$: the NP coupling to quarks is to good accuracy pseudoscalar. In the second case, a good fit is obtained for left-handed NP: $C_{V_L}\ne
0$.

In terms of simplified renormalisable models,
 the $R(D^{(*)})$ anomaly can be solved by the tree-level exchange of a new charged scalar particle \cite{Crivellin:2012ye,Celis:2012dk,Crivellin:2013wna}, as arises in two Higgs doublet models, a heavy charged vector resonance $W'$ \cite{Greljo:2015mma}, or a scalar or vector leptoquark \cite{Fajfer:2015ycq,Becirevic:2016yqi}. A charged Higgs contribution would manifest itself via $C_{S_L},C_{S,R}\ne 0$, while the exchange of a  $W'$ with left-handed couplings generates $C_{V_L} \ne 0$. In leptoquark models, different combinations of Wilson coefficients can be generated, depending on the spin and gauge quantum numbers of the assumed leptoquark.

NP explanations of the $R(D^{(*)})$ anomaly face stringent constraints by complementary measurements in the flavour sector, but also by high-$p_T$ observables. The scalar solution $C_{S_L} \simeq -C_{S_R}$ is challenged by the total $B_c$ lifetime \cite{Li:2016vvp,Alonso:2016oyd,Celis:2016azn}: The large pseudoscalar contribution required to reconcile $R(D^*)$ with the data generates a dangerous contribution to the $B_c\to\tau\nu$ decay, as (pseudo)scalar contributions in the latter mode are not chirality-suppressed. Scalar contributions to the $b\to c\tau\nu$ transition \cite{Celis:2016azn} also modify the $B\to D^{(*)}\tau\nu$ differential decay rates with respect to the SM. While the experimental precision is so far limited, the good agreement of the $B\to D\tau\nu$ differential rate with the SM prediction \cite{Lees:2013uzd,Huschle:2015rga} provides another hint against scalar contributions as the origin of the $R(D^{(*)})$ anomaly. 

NP in $C_{V_L}$, on the other hand, is not subject to the above constraints. Its contribution on the $B_c\to\tau\nu$ decay rate receives the same chirality suppression factor $m_\tau^2/m_b^2$ as in the SM and is therefore safely small. Further, the differential decay rates remain the same as in the SM, since only the overall normalisation changes with the size of $C_{V_L}$. Yet it has been shown \cite{Feruglio:2016gvd,Feruglio:2017rjo} that loop diagrams involving  $C_{V_L}$  generate dangerously large deviations from the SM in $Z$ and $\tau$ decays.

As a result, leptoquark models provide the best NP explanation for the $R(D^{(*)})$ anomaly.
However, also these models are severely constrained. The  $(\bar c b)(\bar \tau\nu)$ operators  are related by the $SU(2)_L$ gauge symmetry to the $(\bar bb)(\bar \tau\tau)$, $(\bar cc)(\bar \tau\tau)$, $(\bar sb)(\bar\tau\tau)$, and $(\bar sb)(\bar\nu\nu)$ operators. The latter two are constrained by the experimental upper bounds on the branching ratios of $B_s\to\tau^+\tau^-$ and $B\to K^{(*)}\nu\bar\nu$ \cite{Calibbi:2015kma,Crivellin:2017zlb}. The former two, on the other hand, are a result of the CKM mixing and lead to deviations from the SM in $\tau$ pair production at the LHC \cite{Faroughy:2016osc}, with the current data excluding large regions of the parameter space. The same interactions also mediate the decays $\Upsilon\to\tau^+\tau^-$ and $\psi\to\tau^+\tau^-$ \cite{Aloni:2017eny}.

Altogether we thus find that a NP resolution of the $R(D^{(*)})$ anomaly is difficult in view of the stringent complementary constraints from $B$ decay observables, but also from EW precision measurements and high-$p_T$ searches. This is not unexpected, given that the required NP contribution is, as discussed earlier, rather large.

\section{\boldmath Rare $b\to s$ transitions and lepton flavour universality}

Recently another set of $B$ decay anomalies has created a lot of excitement, those are related to the semileptonic $b\to s\mu^+\mu^-$ transition. 
While the early hints for a non-standard forward-backward asymmetry $A_\text{FB}$ \cite{Wei:2009zv} were not confirmed by LHCb, the latter experiment found  a $3.7\sigma$ deviation from the SM in the angular observable $P'_5$ of the decay $B\to K^*\mu^+\mu^-$ \cite{Aaij:2013qta}. This anomaly has more recently been confirmed with more statistics by LHCb \cite{Aaij:2015oid}, while the precision achieved at Belle \cite{Wehle:2016yoi}, CMS \cite{CMS:2017ivg}, and ATLAS \cite{ATLAS:2017dlm} is still too low to draw definite conclusions. In addition, a departure from the SM has been found in the differential branching fraction of $B_s\to\phi\mu^+\mu^-$ \cite{Aaij:2015esa}.
However, the substantial departures from unity found in the LFU ratios $R_K$ \cite{Aaij:2014ora} and $R_{K^*}$ \cite{Aaij:2017vbb}, with
\begin{equation}
\label{eq:RK}
R_{K^{(*)}} = \frac{\mathcal{B}(B\to K^{(*)}\mu^+\mu^-)}{\mathcal{B}(B\to K^{(*)}e^+e^-) }\,,
\end{equation}
 are even more intriguing, as those ratios are theoretically extremely clean.

The semileptonic $b\to s\mu^+\mu^-$ and radiative $b\to s\gamma$ transitions are theoretically described by the effective Hamiltonian
\begin{equation}
\mathcal{H}_\text{eff}= -\frac{4 G_F}{\sqrt{2}} V_{tb}^* V_{ts} \frac{e^2}{16\pi^2}\sum_i(C_i {\cal O_i} +C'_i {\cal O'_i})+h.c.\,,
\end{equation}
where the operators most sensitive to NP are the dipole operators 
\begin{equation}\label{eq:C7}
\mathcal{O}^{(\prime)}_7 =\frac{m_b}{e} (\bar s \sigma_{\mu\nu} P_{R(L)}b) F^{\mu\nu}
\end{equation}
and the four fermion operators 
\begin{equation}\label{eq:C910}
\mathcal{O}^{(\prime)}_9 = (\bar s\gamma_\mu P_{L(R)} b)(\bar\mu\gamma^\mu\mu)\,,\qquad
\mathcal{O}^{(\prime)}_{10}  = (\bar s\gamma_\mu P_{L(R)} b)(\bar\mu\gamma^\mu\gamma_5\mu)\,,
\end{equation}
as they are not affected by tree-level contributions in the SM, making a sizeable NP effect much easier to achieve. For the sake of simplicity we neglect the scalar and pseudoscalar operators $\mathcal{O}^{(\prime)}_{S,P}$  as they are strongly constrained by the measured $B_s\to\mu^+\mu^-$ branching ratio \cite{Aaij:2017vad}, agreeing well with its SM prediction \cite{Bobeth:2013uxa}.

The measurements of various observables in radiative and semileptonic $b\to s$ transitions constrain the values of the Wilson coefficients $C^{(\prime)}_{7,9,10}$. For example, the radiative decays $B\to X_s\gamma$, $B\to K^*\gamma$ etc.\ are driven only by the magnetic dipole operators $\mathcal{O}^{(\prime)}_7$, while the semileptonic decays $B\to K^{(*)}\mu^+\mu^-$, $B\to X_s\mu^+\mu^-$, and $B_s\to\phi\mu^+\mu^-$ are sensitive to all six Wilson coefficients. Further information can be obtained from the study of the full angular distribution of the latter decays. It can thus be tested whether anomalies in various observables have a consistent NP interpretation.

\begin{figure}
\centering{\includegraphics[width=.48\textwidth]{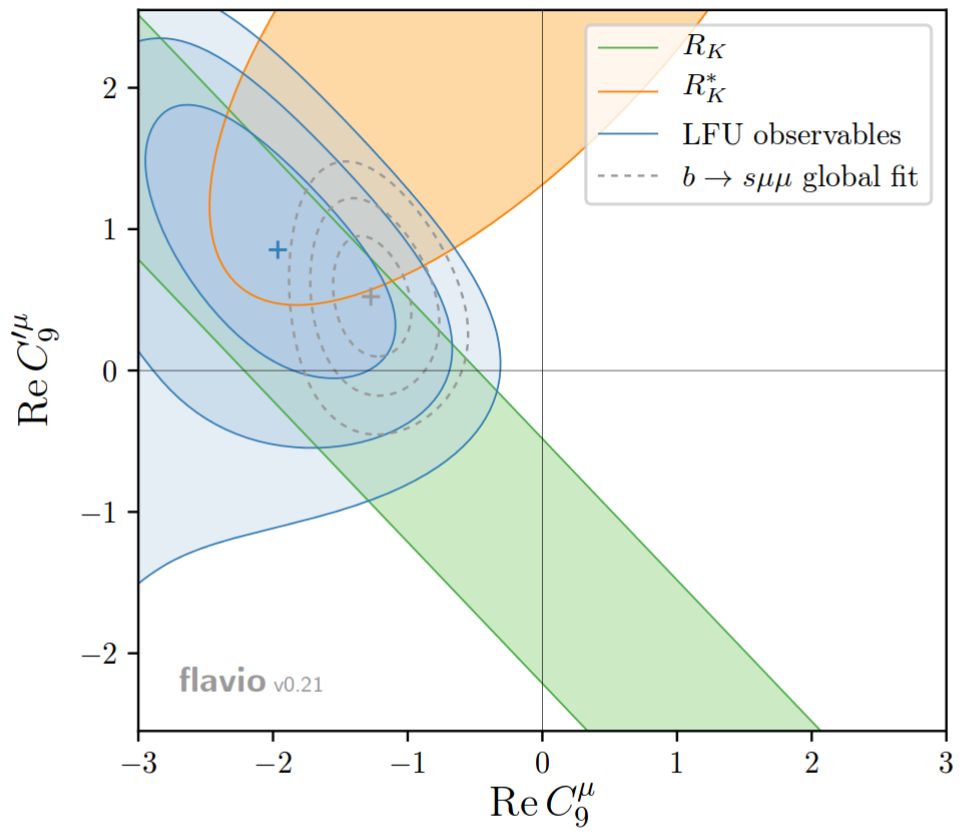}\quad
\includegraphics[width=.41\textwidth]{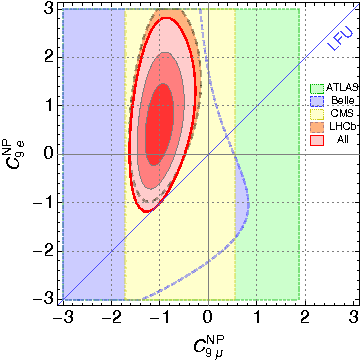}}
\caption{{\it left:} Two-dimensional fit allowing for NP in $C_9$ and $C'_9$. Figure taken from \cite{Altmannshofer:2017yso}.
{\it right:} Two-dimensional fit allowing for independent NP contributions in $C_{9\mu}$ and $C_{9e}$, showing separately the input from various experiments. Figure taken from \cite{Capdevila:2017bsm}.\label{fig:C9fit}}
\end{figure}

Several groups \cite{Altmannshofer:2017yso,Capdevila:2017bsm,DAmico:2017mtc,Geng:2017svp} have performed global fits of the relevant Wilson coefficients to the data, with the result that a quite large NP contribution to the Wilson coefficient $C_9$
 is required to significantly ($>4\sigma$) improve  the goodness of fit with respect to the SM, as can be seen from the left panel of figure \ref{fig:C9fit}. The fit also allows for non-negligible NP contributions to $C'_9$ and/or $C_{10}$, however the latter are not as strictly required. The right panel of figure \ref{fig:C9fit} shows that the discrepancy with the SM prediction is mainly driven by the LHCb data, calling for an independent confirmation by the other experimental collaborations. 

A comment is in order concerning hadronic uncertainties in the semileptonic decays $B\to K^{(*)}\mu^+\mu^-$. These decays are well described in terms of $B\to M$ form factors that contain the non-perturbative interactions between the decaying $B$ meson and the final state meson $M=K,K^*$. These form factors can be determined by lattice QCD \cite{Horgan:2013hoa} and light-cone sum rule \cite{Straub:2015ica} techniques, allowing for further systematic improvements in the near future. The non-factorisable corrections \cite{Khodjamirian:2010vf,Jager:2012uw,Descotes-Genon:2014uoa,Jager:2014rwa,Lyon:2014hpa}, on the other hand, cannot be treated systematically, so that their size can only be estimated. The dominant contribution in this context stems from charm loop effects that should be most relevant in the $q^2$ region below the $c\bar c$ resonance threshold.

In order to avoid the uncertainties from hadronic effects, various observables have been constructed in which those uncertainties cancel. The observables $P_i, P'_i$ describing the angular distribution of the $B\to K^*\mu^+\mu^-$ final state \cite{Matias:2012xw,DescotesGenon:2012zf} have been designed such that they are form-factor independent at leading order.  Yet they are still sensitive to non-factorisable effects \cite{Jager:2012uw,Jager:2014rwa}. In order to get rid of the latter uncertainties, the LFU ratios $R_K$, $R_{K^*}$ defined in \eqref{eq:RK} have been proposed \cite{Hiller:2014ula}. These are theoretically extremely clean \cite{Bordone:2016gaq}, as in the SM the only departure from unity is generated by the very small muon mass. Hence,
the anomalies at the $2.5\sigma$ level reported by LHCb in the LFU ratios $R_K$ and $R_{K^*}$ are particularly striking. If these anomalies are independently confirmed by other measurements, the presence of LFU violating NP will unambiguously be established. At the moment, however, we have to be patient and wait for further experimental investigation, like measurements of additional LFU observables and independent confirmations by other experimental collaborations. Fortunately, the LHC experiments are currently collecting more data, and the first physics run of Belle 2 will start in late 2018, so we are confident to have a definite answer soon.

Interestingly, the global fits discussed above \cite{Altmannshofer:2017yso,Capdevila:2017bsm,DAmico:2017mtc,Geng:2017svp} show that the anomalies in $R_K$ and $R_{K^*}$ can be resolved by the same NP contribution as the $P'_5$ anomaly, if the NP is assumed to contribute only to the muon channel. Yet, due to the sizeable uncertainties,  also a significant NP effect in the electron channel is possible at present. Note that the strong suppression of $R_{K^*}$ below the SM prediction in the region $q^2 \in [0.045,1.1]\,\text{GeV}^2$ calls for a lepton flavour dependent NP contribution to $C_{10}$ \cite{Geng:2017svp}. It is however impossible to accommodate the experimental central value by means of NP.

After identifying the necessary NP pattern in the effective theory language, let us now consider possible NP models that generate the required contributions. In the most widely discussed NP models, the $b\to s\mu^+\mu^-$ transition is mediated at the tree level, thereby providing a good explanation of the relatively large NP contribution to $C_9$, whose SM contribution is loop-suppressed. Particularly popular are models with  an extra neutral $Z'$ gauge boson mediating the flavour changing $b\to s$ transitions, and coupling to muons \cite{Altmannshofer:2013foa,Gauld:2013qba,Buras:2013qja,Buras:2013dea,Crivellin:2015lwa,Chiang:2016qov,DiChiara:2017cjq}. For instance, a model with gauged $L_\mu-L_\tau$ symmetry has been suggested in \cite{Altmannshofer:2014cfa} and subsequently studied in \cite{Crivellin:2015mga,Altmannshofer:2015mqa,Altmannshofer:2016oaq,Altmannshofer:2016jzy}. The possibility of a $Z'$  resonance of a composite sector has also been investigated \cite{Niehoff:2015bfa,DAmico:2017mtc}. Typically however, in the latter class of models, a different pattern of NP effects arises \cite{Altmannshofer:2013foa}. The same conclusion had also been drawn earlier in the context of Randall-Sundrum models \cite{Blanke:2008yr,Bauer:2009cf,Blanke:2012tv}, being dual to a certain type of composite models.

Leptoquark models \cite{Hiller:2014yaa,Gripaios:2014tna,Dorsner:2016wpm} consitute another popular explanation of the observed anomaly. Like in the case of the $R(D^{(*)})$ anomaly, also here several realisations in terms of the leptoquark spin and gauge representation are possible. Note that in this type of models, large LFU violating effects are particularly easy to accommodate.

It is also possible to address the $b\to s$ anomalies by loop-induced NP contributions. New box contributions \cite{Gripaios:2015gra,Arnan:2016cpy} and $Z'$ penguins \cite{Kamenik:2017tnu} have been discussed in the literature. A $Z'$ model with a loop-induced coupling to muons has been investigated in \cite{Belanger:2015nma}. On the contrary, models inducing $Z$ penguin effects, like the MSSM \cite{Altmannshofer:2013foa} or the Littlest Higgs model with T-parity \cite{Blanke:2006eb,Blanke:2015wba}, are not compatible with the requirement of a large NP contribution to the Wilson coefficient $C_9$ and do not generate new LFU violating effects.

Models explaining the $b\to s$ anomalies are generally constrained by the well-measured $B_s-\bar B_s$ mixing observables \cite{Altmannshofer:2013foa}, the experimental upper bounds on the $B\to K^{(*)}\nu\bar\nu$ decay rates \cite{Buras:2014fpa,Calibbi:2015kma}, as well as the anomalous magnetic moment of the muon, $(g-2)_\mu$ \cite{Bauer:2015knc}. However, in this case an agreement with the data is easier to achieve than in the case of the $R(D^{(*)})$ anomaly. Additionally, also high-$p_T$ data play a role in constraining possible NP models -- this time by the measured SM-like high-$p_T$ dilepton tails \cite{Greljo:2017vvb}.

\section{Outlook}

In this contribution we have presented an overview of the current most exciting anomalies in flavour physics. While it remains to be seen which of them are indeed caused by NP, their study provides interesting insights on the structure of possible extensions of the SM. 

If eventually all of the discussed anomalies will turn out to have a NP origin, the challenge will be to identify their common NP origin. For the $B$ physics anomalies, a model building guide for a combined explanation has recently been provided \cite{Buttazzo:2017ixm}. However, the models discussed as solutions of the $\epe$ anomaly are largely complementary to the ones employed in the $B$ sector. 

Further input on the NP at work can be expected from future measurements in the flavour sector. Additional $K$ and $B$ meson decay observables, as discussed in the text, will play an important role here. Furthermore, charged lepton flavour violating effects are intimately linked to LFU violation, and a near-future detection can thus be hoped for. An exciting era of NP discoveries in the flavour sector may  just have begun. 

\paragraph{Acknowledgements}
I would like to thank the organizers of FPCP2017 for putting together such a stimulating and enjoyable  conference, and for giving me the opportunity to present this overview.

\end{document}